\documentclass[prb,aps,eqsecnum]{revtex4}% Physical Review B
\usepackage{graphicx}
\newcommand{\e}{{\rm e}}
\newcommand{\ep}{\varepsilon}
\newcommand{\tr}{{\rm Tr}}
\newcommand{\be}{\begin{equation}}
\newcommand{\ee}{\end{equation}}
\newcommand{\ba}{\begin{eqnarray}}
\newcommand{\ea}{\end{eqnarray}}
\begin{document}
%\tightenlines
\title{Quantum Statistical Calculations and \\
Symplectic Corrector Algorithms}

\author{Siu A. Chin}

\affiliation{Department of Physics, Texas A\&M University,
College Station, TX 77843, USA}

%\date{\today}
\begin{abstract}
%%
%%%%%%%%%%%%%%%%%%%%%%%%%%%%%%%%%%%%%%%%%%%%%%%%%%%%%%%%%%%%%%%%%%%%%%%%
%%
%%  Abstract
%%
%%%%%%%%%%%%%%%%%%%%%%%%%%%%%%%%%%%%%%%%%%%%%%%%%%%%%%%%%%%%%%%%%%%%%%%%
%%
The quantum partition function at finite temperature requires computing
the trace of the imaginary time propagator. For numerical
and Monte Carlo calculations, the propagator is
usually split into its kinetic and potential parts. A higher
order splitting will result in a higher order convergent algorithm.
At imaginary time, the kinetic energy propagator is usually the
diffusion Greens function. Since diffusion cannot be simulated backward in
time, the splitting must maintain the positivity of all intermediate time
steps. However, since the trace is invariant under similarity transformations
of the propagator, one can use this freedom to ``correct"
the split propagator to higher order. This use of similarity transforms
classically give rises to symplectic corrector algorithms. The
split propagator is the symplectic kernel and the similarity
transformation is the corrector. This work proves a generalization of
the Sheng-Suzuki theorem: no positive time step propagators with only
kinetic and potential operators can be corrected beyond second order.
Second order forward propagators can have fourth order traces
only with the inclusion of an additional commutator. We give detailed
derivations of four forward correctable second order
propagators and their minimal correctors.

\end{abstract}
\maketitle

\section {Introduction}
The quantum partition function requires computing the
trace
\be
Z=\tr(\rho)=\tr( \e^{-\beta H} ),
\label{part}
\ee
where $\rho$ is imaginary time propagator, $\beta=1/(k_B T)$ is the
inverse temperature and $H=T+V$ is the usual
Hamiltonian operator.
Although specific forms of the kinetic and potential energy
operators will not be used in the following, it is useful to keep
in mind the many-body case where
$T=(-\hbar^2/2m)\sum_{i=1} \nabla^2_i$ and $V=\sum_{i<j}v(r_{ij})$.
In numerical or Monte Carlo calculations, the imaginary time propagator
is first discretized as
\be
 \e^{-\beta(T+V)}=\left[\e^{\ep (T+V)}\right]^n,
\ee
where $\ep=-{\mit\Delta}\beta=-\beta/n$, and the short-time
propagator $\e^{\ep (T+V)}$ is then approximated in various ways.
One systematic method is to decompose, or split the short-time
propagator into the product form
\be {\rm e}^{\ep (T+V
)}\approx\prod_{i=1}^N {\rm e}^{t_i\ep T}{\rm e}^{v_i\ep V},
\label{arb}
\ee
with coefficients $\{t_i, v_i\}$ determined by the
required order of accuracy. For quantum statistical calculations,
since $\langle{\bf r}^\prime|\,{\rm e}^{t_i\ep T}|\,{\bf r}\rangle
\propto {\rm e}^{-({\bf r}^\prime-{\bf
r})^2/(2t_i{\mit\Delta}\beta)}$ is the diffusion kernel, the
coefficient $t_i$ must be positive in order for it to be simulated
or integrated. If $t_i$ were negative, the kernel is unbounded and
unnormalizable, and no probabilistic based (Monte Carlo)
simulation is possible. However, as first proved by
Sheng\cite{sheng}, and later by Suzuki\cite{suzukinogo}, beyond
second order, any factorization of the form (\ref{arb}) {\it must}
contain some negative coefficients in the set $\{t_i, v_i\}$.
Goldman and Kaper\cite{goldman} further proved that any
factorization of the form (\ref{arb}) must contain at least one
negative coefficient for {\it both} operators. Thus, despite
myriad of factorization schemes of the form (\ref{arb}) proposed
in the classical symplectic integrator
literature\cite{mcl91,mcl95,kos96,mcl02}, none can be used for
doing quantum statistical calculations beyond second order. It is
only recently that fourth order, all positive-coefficient
factorization schemes have been found\cite{suzfour,chin} and
applied to time-irreversible problems containing the diffusion
kernel\cite{fchinl,fchinm,auer,ochin,jang}. In order to bypass the
Sheng-Suzuki's theorem, one must include other operators in the
factorization (\ref{arb}), such as the double commutator
$[V,[T,V]]$, where $[A,B]\equiv AB-BA$.

In computing the quantum partition function $Z$, only the trace of
$\rho=\e^{-\beta H}$ is required. Since the trace is invariant
under the similarity transformation \be \tilde\rho=S\rho S^{-1},
\label{stran} \ee one is free to use any such $\tilde\rho$ to
compute $Z$. This is immaterial if $\rho$ is known exactly.
However, if the short-time propagator is only known approximately,
then one may use a clever choice of $S$ to further improve the
approximation. This is a well known idea in many areas of physics.
For example, to calculate the exact quantum many-body ground state
using the Diffusion Monte Carlo algorithm, one can choose
$S=\phi_0$, where $\phi_0$ is a known trial function close to the
exact ground state. This is the idea of ``importance sampling" as
introduced by Kalos {\it et al.}\cite{kalos}. Its operator
formulation as described above has been implemented by
Chin\cite{chin90} some time ago. Similar ideas have been used to
improve path-integrals, as detailed by Kleinert\cite{klein}. If
the short time propagator is approximated by the product form
(\ref{arb}), the error terms can be calculated explicitly and
eliminated by $S$. When implemented classically, these are known
as symplectic ``corrector", or ``process"
algorithms\cite{wis96,mcl962,mar96,mar97,blan99,blan01}. In this
context the propagator $\rho$ is the kernel algorithm and $S$ is
the corrector. Since $S$ disappears in the calculation of $Z$,
there is no restriction on the form of $S$. If $S$ were also
expanded in the product form (\ref{arb}), there is no restriction
on the sign of its coefficients. This suggests that there may
exist a product form (\ref{arb}) of $\rho$ with only positive
coefficients such that its trace is correct to higher order. This
would not be precluded by the existing Sheng-Suzuki theorem.

In this work, we show that this is not possible. If $\rho$
is approximated by the product form (\ref{arb}) with positive
coefficients $\{t_i\}$, then $\tilde\rho$ cannot be corrected
by $S$ to higher than second order. The proof of this
generalizes the Sheng-Suzuki theorem.
The corrected propagator $\tilde\rho$ can
be fourth order only if additional operators, such as
$[V,[T,V]]$, are used in the splitting of $\rho$. By understanding
the ``correctability" requirement, we can systematically deduce
the four fundamental correctable second order propagators
and their correctors.

In the following Section, we recall some basic results of similarity
transforms. Beyond second order, only a special class of
approximate $\rho$ satisfying the ``correctability" condition can be
corrected to higher order. In Section III, we compute the explicit form
of the error coefficients required by the correctability
criterion. In Section IV, we show that this requirement cannot be
satisfied for propagators of the product form (\ref{arb}) with
only positive $\{t_i\}$ coefficients. In Section V, based on our
understanding of the correctability restriction, we deduce all four
second order correctable propagators and their minimal correctors.
Some conclusions are given in Section VI.

\section {Similarity Transforms and the Correctability Criterion}

Similarity transforms on approximate propagators of the product
form (\ref{arb}) have been studied extensively in the context of
symplectic correctors\cite{wis96,mcl962,mar96,mar97,blan99}.
However, not all use the language of operators and some are
specific to celestial mechanics. Here, we recall some elementary
results and establish the fundamental correctability requirement
in the context of quantum statistical physics.

Since \be S\rho S^{-1}=\left[S\e^{\ep (T+V)}S^{-1}\right]^n, \ee
it is sufficient to study the similarity transformation of the
approximate short-time propagator $\rho_A$. Let $\rho_A$
approximates $\e^{\ep (T+V)}$ in the product form such that
\be
\rho_A=\prod_{i=1}^N {\rm e}^{t_i\ep T}{\rm e}^{v_i\ep V}=\e^{\ep
H_A}, \label{arho}
\ee
where $H_A$ is the approximate Hamiltonian
\be H_A=T+V+\ep (e_{TV}[T,V])+\ep^2(\,
e_{TTV}[T,[T,V]]+e_{VTV}[V,[T,V]]\,)+O(\ep^3) \label{ha}
\ee
with error coefficients $e_{TV}$, $e_{TTV}$, $e_{VTV}$ determined by
factorization coefficients $\{t_i,v_i\}$. The transformed
propagator is
\be \tilde\rho_A=S\rho_A S^{-1}=S\e^{\ep H_A}S^{-1}
=\e^{\ep( S H_A S^{-1})}=\e^{\ep\widetilde H_A},
\label{tha}
\ee
where the last equality defines the transformed approximate
Hamiltonian $\widetilde H_A$. If now we take
\be S=\exp[\ep C]
\label{scor}
\ee
where $C$ is the to-be-determined corrector, then
we have the fundamental result
\be \widetilde H_A=e^{\ep
C}H_A\e^{-\ep C}=H_A+\ep[C,H_A] +{1\over
2}\ep^2[C,[C,H_A]]+{1\over{3!}}\ep^3[C,[C,[C,H_A]]]+\cdots .
\label{fcor} \ee

Let's first consider the case where the product form (\ref{arho}) for
$H_A$ is left-right symmetric, {\it i.e.}, either $t_1=0$ and
$v_i=v_{N-i+1}$, $t_{i+1}=t_{N-i+1}$, or $v_N=0$ and
$v_i=v_{N-i}$, $t_{i}=t_{N-i+1}$. In this cases, the propagator
is reversible, $\rho_A(\ep)\rho_A(-\ep)=1$, and $H_A(\ep)$ is an even
function of $\ep$ with $e_{TV}=0$.
In this case
\ba
\widetilde H_A&=&H_A+\ep[C,H_A] +\cdots,\nonumber\\
&=& T+V+\ep^2(\, e_{TTV}[T,[T,V]]+e_{VTV}[V,[T,V]]\,)+\ep[C,T+V]+ \cdots,
\label{gcor}
\ea
and one immediately sees that the choice $C=\ep C_1$ with
$C_1\equiv c_{TV}[T,V]$ would eliminate either second order error
term with $c_{TV}=e_{TTV}$ or $c_{TV}=e_{VTV}$. So, if $H_A$ is
constructed such that
\be
e_{TTV}=e_{VTV}
\label{eqcof}
\ee
then {\it both} can be simultaneously eliminated by the corrector.
This is the fundamental ``correctability" requirement for correcting
a second order $\rho_A$ to fourth order.
This observation can be generalized to higher order.
At higher orders, $H_A$ will have error terms of the form
$[T,Q_i]$ and $[V,Q_i]$ where $Q_i$ are some higher order commutator
generated by $T$ and $V$. If $H_A$ is of order $2n$ in $\ep$,
then $\widetilde H_A$ can be of order $2n+2$ only if $H_A$'s error coefficients
for $[T,Q_i]$ and $[V,Q_i]$ are {\it equal} for all $Q_i$'s. This fundamental
corrector insight is often obscured by the more general case where
odd order errors are allowed.

Sheng\cite{sheng} and Suzuki\cite{suzukinogo} independently proved
that no $\rho_A$ of the form (\ref{arho}) can have positive
coefficients $t_i$ beyond second order. More precisely, if $\rho_A$ is
of the product form (\ref{arho}) with positive $t_i$'s such that $e_{TV}=0$,
then $e_{TTV}$ and $e_{VTV}$ cannot both be zero. We will prove a more
general theorem that the product form (\ref{arho}) with positive $t_i$'s
such that $e_{TV}=0$ cannot be {\it corrected} beyond second order,
{\it i.e.,} $e_{TTV}$ can never {\it equal} to $e_{VTV}$. From this
perspective, the Sheng-Suzuki theorem is a special case
where the common value for both coefficients is zero.

In the general case where $e_{TV}\neq 0$, we have
\ba
\widetilde H_A&&=T+V+\ep (e_{TV}[T,V])
+\ep^2(\, e_{TTV}[T,[T,V]]+e_{VTV}[V,[T,V]]\,)\nonumber\\
&&+\ep[C,T+V] +\ep^2e_{TV}[C,[T,V]]+{1\over 2}\ep^2[C,[C,T+V]]
+O(\ep^3).
\label{thatv}
\ea
Since $[c_T T+ c_V V,T+V]=(c_T-c_V)[T,V]$,
the liner term in $\ep$ can be eliminated if we choose
$C=C_0\equiv c_T T+ c_V V$ such that
\be (c_T-c_V)=-e_{TV}.
\label{first}
\ee
This is the first order correctability condition.
This means that with a suitable choice of $c_T$ and $c_V$, a
first order propagator can always be corrected to
second order. Hence, {\it the trace of any first order propagator
is always second order}. For example, the trace
$\tr(\e^{\ep T}e^{\ep V})$ is second order despite its appearance.

With the first order correctability condition satisfied, the
remaining commutators in (\ref{thatv}) are either $[T,[T,V]]$ or
$[V,[T,V]]$, and can again be corrected by adding to $C$ the term
$\ep C_1=\ep c_{TV} [T,V]$. Thus with \be C=C_0+\ep C_1=c_T T+ c_V
V+\ep c_{TV} [T,V] \label{corform} \ee such that
$(c_T-c_V)=-e_{TV}$, we have \ba \widetilde H_A&&= T+V
+\ep^2(\, e_{TTV}[T,[T,V]]+e_{VTV}[V,[T,V]]\,)\nonumber\\
&&+\ep^2[C_1,T+V] +\ep^2e_{TV}[C_0,[T,V]]+{1\over 2}\ep^2[C_0,[C_0,T+V]]
+O(\ep^3),\nonumber\\
&&=T+V
+\ep^2(e_{TTV}-c_{TV}+{1\over 2}c_T e_{TV})[T,[T,V]]\nonumber\\
&&\qquad\qquad + \ep^2(e_{VTV}-c_{TV}+{1\over 2}c_V e_{TV})[V,[T,V]] +O(\ep^3)
\label{thado}
\ea
If we now choose $c_{TV}=e_{TTV}+{1\over 2}c_T e_{TV}$ to eliminate
the error term $[T,[T,V]]$, then the error term $[V,[T,V]]$ can
vanish only if
\be
e_{TTV}=e_{VTV}+{1\over 2}(e_{TV})^2.
\label{cortwo}
\ee
This is the general second order correctability requirement
for correcting any first order propagator beyond second order.
The major result of this work is to show that this condition
cannot be satisfied
for product decomposition of the form (\ref{arho}) with
only positive $t_i$ coefficients.

\section{Determining the Error Coefficients}

To check whether the correctability requirement (\ref{cortwo}) can ever
be satisfied by an approximate propagator of the product
form (\ref{arho}), we need to determine
$e_{TV}$, $e_{TTV}$ and $e_{VTV}$ in terms of $\{t_i,v_i\}$. From the
assumed equality
\be
\prod_{i=1}^N
{\rm e}^{t_i\ep T}{\rm e}^{v_i\ep V}=\e^{\ep H_A},
\label{arhot}
\ee
with $H_A$ given by (\ref{arho}), we can expand both sides and compare
terms order by order in powers of $\ep$.
The left hand side
of (\ref{arhot}) can be expanded as
\be
{\rm e}^{\ep t_1 T}{\rm e}^{\ep v_1 V}
{\rm e}^{\ep t_2 T}{\rm e}^{\ep v_2 V}
\cdots
{\rm e}^{\ep t_N T}{\rm e}^{\ep v_N V}
=1+\ep \left ( \sum_{i=1}^N t_i\right ) T
+\ep \left( \sum_{i=1}^N v_i\right) V +\cdots,
\label{rhoalh}
\ee
and the right hand side as
\ba
\e^{\ep H_A}&&=1+\ep(T+V)+
\ep^2 e_{TV}[T,V]+\ep^3e_{TTV}[T,[T,V]]+\ep^3 e_{VTV}[V,[T,V]]\nonumber\\
&&\qquad +{1\over 2}\ep^2(T+V)^2+{1\over 2}\ep^3e_{TV}
      \left \{(T+V)[T,V]+[T,V](T+V)\right \}\nonumber\\
&&\qquad +{1\over {3!}}\ep^3(T+V)^3+\cdots \label{rhoarh}
\ea
Matching the first order terms in $\ep$ gives the primary
constraints
\be \sum_{i=1}^N t_i=1\quad{\rm and}\quad \sum_{i=1}^N
v_i=1. \label{tvcon}
\ee To determine the error coefficients, we
``tag" a particular operator in (\ref{rhoarh}) whose coefficient
contains $e_{TV}$, $e_{TTV}$ or $e_{VTV}$ and match the same
operator's coefficients in the expansion of (\ref{rhoalh}). For
example, in the $\ep^2$ terms of (\ref{rhoarh}), the coefficient
of the operator $TV$ is $({1\over 2}+e_{TV})$ Equating this to the
coefficients of $TV$ from (\ref{rhoalh}) gives
\be {1\over
2}+e_{TV}=\sum_{i=1}^N s_i v_i. \label{etv} \ee
where we have
introduced the variable
\be s_i=\sum_{j=1}^i t_j. \label{si} \ee
This way of computing $TV$ from (\ref{rhoalh}) corresponds to
first picking out a $V$ operator from among all the $v_i$ terms,
then combine all the $t_i$ terms to its left in the exponential to
generate a $T$ operator. Alternatively, the same coefficient can
also be expressed as
\be {1\over 2}+e_{VT}=\sum_{i=1}^N t_i u_i.
\label{evt}
\ee where
\be u_i=\sum_{j=i}^N v_j. \label{ui}
\ee
This way of computing $TV$ corresponds to first picking out a $T$
operator from among all the $t_i$ terms, then combine all the
$v_i$ terms to its right in the exponential to generate a $V$
operator. To demonstrate how these variables are to be used, we
can directly prove the equality of (\ref{etv}) and (\ref{evt}).
First, note that $s_N=1$ and $u_1=1$. Second, since
$t_i=s_i-s_{i-1}$, at $i=1$ we must consistently set $s_0=0$.
Similarly, since $v_i=u_i-u_{i+1}$, we must set $u_{N+1}=0$.
Therefore we have
\be \sum_{i=1}^N s_i v_i=\sum_{i=1}^N s_i
(u_i-u_{i+1}) =\sum_{i=1}^N (s_i-s_{i-1}) u_i=\sum_{i=1}^N t_i u_i
\label{ptv} \ee

The determination of error coefficients is simplified if we pick
operators whose expansion coefficients are easy to calculate.
Matching the coefficients of operators $TTV$ and $TVV$ (note, {\it
not} the operator $VTV$) yields \ba {1\over 6}+{1\over
2}e_{TV}+e_{TTV}&&={1\over 2}\sum_{i=1}^N s_i^2 v_i ={1\over
2}\sum_{i=1}^N (s_i^2-s^2_{i-1}) u_i,
\label{ettv}\\
{1\over 6}+{1\over 2}e_{TV}-e_{TVT}&&={1\over 2}\sum_{i=1}^N t_i u_i^2.
\label{evtv}
\ea

\section{Proving the Main Result}

Using the expression for $e_{TVT}$ from (\ref{evtv}), the
correctability requirement (\ref{cortwo}) reads \be {1\over
2}\sum_{i=1}^N t_i u_i^2=a, \label{corthr} \ee with \be a={1\over
2}({1\over 2}+e_{TV})^2+{1\over{24}}-e_{TTV} \label{acof} \ee and
$e_{TV}$, $e_{TTV}$ given by (\ref{evt}), (\ref{ettv})
respectively. In Suzuki's proof\cite{suzukinogo}, he recognizes
that in terms of the variable $\sqrt{t_i}u_i$,  (\ref{corthr}) is
a hypersphere and (\ref{evt}), (\ref{ettv}) are hyperplanes. His
proof is based on a geometric demonstration that his hyperplane
cannot intersect his hypersphere. While this geometric language is
very appealing, it is cumbersome when dealing with more than one
hyperplane. We will use a different strategy.

If $t_i$ are all positive, then the LHS of (\ref{corthr}) is a
positive-definite quadratic form in $u_i$. There would be no real
solutions for $u_i$ if the minimum of the quadratic form is
greater than $a$. Our strategy is therefore to minimize the
quadratic form subject to constraints (\ref{evt}) and (\ref{ettv})
\ba \sum_{i=1}^N t_i u_i&=&b,
\label{cb}\\
\sum_{i=1}^N t_i(s_i+s_{i-1}) u_i&=&c, \label{cc} \ea with
$b={1\over 2}+e_{VT}$, $c={1\over 3}+e_{TV}+2e_{TTV}$, and show
that the resulting minimum is always greater than $a$. (The
primary constraints (\ref{tvcon}) are just $s_N=1$ and $u_1=1$.)

For constrained minimization, one can use the method of Lagrange
multiplier. Minimizing \be F = {1\over 2}\sum_{i=1}^N t_i u_i^2
-\lambda_1 \left( \sum_{i=1}^N t_i u_i-b\right) -\lambda_2 \left(
\sum_{i=1}^N t_i(s_i+s_{i-1}) u_i-c\right ) \ee gives \be
u_i=\lambda_1+\lambda_2(s_i+s_{i-1}). \label{usol} \ee
Substituting this back to satisfy constraints (\ref{cb}) and
(\ref{cc}) determines $\lambda_1$ and $\lambda_2$: \be
\lambda_1+\lambda_2=b, \ee \be \lambda_1+\lambda_2+g\lambda_2=c.
\ee The only non-trivial evaluation is $\sum_{i=1}^N
t_i(s_i+s_{i-1})^2=1+g $, where \be g=\sum_{i=1}^N
(s_i^2s_{i-1}-s_is_{i-1}^2). \ee The minimum of the quadratic form
is therefore \ba
F&=&{1\over 2}\sum_{i=1}^N t_i [\lambda_1+\lambda_2(s_i+s_{i-1})]^2\nonumber\\
 &=&{1\over 2}[ (\lambda_1+\lambda_2)^2+g\lambda_2^2]\nonumber\\
 &=&{1\over 2}[b^2+ {1\over g}(c-b)^2].
\label{min}
\ea
To minimize $F$, one must maximize $g$. Solving $\partial g/\partial s_i=0$
gives $s_i=(s_{i+1}+s_{i-1})/2$,
which means that $s_i$ is linear in $i$.
The normalization $s_N=1$ fixes $s_i=i/N$, giving
\be
g_{max}={1\over 3}(1-{1\over{N^2}}).
\label{gval}
\ee
This is indeed a maximum
since one can directly verify that
$\partial^2 g/\partial s_i^2=-2(s_{i+1}-s_{i-1})<0$.
Hence, at any finite $N$,
\be
F>{1\over 2}[b^2+ 3(c-b)^2]=
{1\over 2}({1\over 2}+e_{TV})^2+{3\over 2}(2\, e_{TTV}-{1\over 6})^2
=a+ 6\,e_{TTV}^2.
\label{fin}
\ee
Thus the minimum of the quadratic form is always higher than the value
required by the correctability condition. Hence, no real solutions
for $u_i$ are possible if $t_i$ are all positive.

We note that the above proof is independent of $e_{TV}$.
For $e_{TV}=0$, the correctability condition is
just $e_{TTV}=e_{VTV}$. Hence for symmetric decompositions
with positive $t_i$'s, where $e_{TV}=0$ is automatic, we have as a
corollary that $e_{TTV}$ can never equal to $e_{VTV}$.

\section{Correctable forward propagators and their correctors}

The last section is the main result of this work. Here, we show
how the correctability criterion can be applied systematically to
deduce forward correctable second order propagators and their minimal
correctors.

The proof of non-correctability is limited to the conventional
product form (\ref{arho}), which factorizes the propagator only in
terms of operators $T$ and $V$. As shown in the last section,
symmetrically decomposed positive-time-step propagators cannot be
corrected beyond second order because $e_{TTV}$ cannot be made
equal to $e_{VTV}$. For example, the second order propagator
\be
\exp({1\over 2}\ep T) \exp( \ep V ) \exp({1\over 2}\ep T)
\label{proptw} \ee
has $t_1=t_2=1/2$, $v_1=u_1=1$, $s_1=1/2$ and
$e_{TV}=0$. From (\ref{ettv}) and (\ref{evtv}), we can determine
indeed that the two error coefficients are not equal:
\ba
e_{TTV}&=&{1\over 2}\left({1\over 2}\right)^2 1-{1\over 6}=-{1\over {24}},\nonumber\\
e_{TVT}&=&{1\over 6}-{1\over 2}\left({1\over 2}\right)1=-{1\over {12}}.
\label{tvtv}
\ea
A simple way to force them equal is to directly incorporate either
operator $[T,[T,V]]$ or $[V,[T,V]]$ in the factorization process.
Since $[V,[T,V]]=(\hbar^2/m)\sum_i |\nabla_i \sum_{j\ne i}v(r_{ij})|^2$
is just another potential function, Suzuki\cite{suz95} suggested that one
should keep the operator $[V,[T,V]]$. If now we add
${1\over {24}}\ep^3[V,[T,V]]$ to $\ep V$ in (\ref{proptw}), we can
change the coefficient $e_{VTV}$ from $-1/12$ to $-1/24$,
matching that of $e_{TTV}$. The result is still only a second order
propagator
\be
\rho_{TI}=
\exp\left ({1\over 2}\ep T\right )
\exp\left ( \ep V + {1\over {24}}\ep^3[V,[T,V]] \right )
\exp\left ({1\over 2}\ep T\right ),
\label{ti}
\ee
but now has a fourth order trace.
This propagator was first obtained by Takahashi and Imada\cite{ti,li}
by directly computing the trace. It is a remarkable find given how
little they had to work with. This derivation explains, without
doing any trace calculation, why the propagator worked.

The alternative of keeping $[T,[T,V]]$ would require adding
$-{1\over {24}}\ep^3[T,[T,V]]$ to make $e_{TTV}$ equal to
$e_{VTV}$'s value of $-1/12$. This operator is too complicated for
practical use, but in the case of the harmonic oscillator, it can
be combined with the kinetic energy operator:
\be
\rho^{\,\prime}_{2B}= \exp\left ({1\over 2}\ep T- {1\over
{48}}\ep^3[T,[T,V]]\right ) \exp\left ( \ep V  \right ) \exp\left
({1\over 2}\ep T- {1\over {48}}\ep^3[T,[T,V]]\right ).
\label{newtbp}
\ee
This can also be written in the form of
\be
\rho_{2B}= \exp\left ({1\over 2} \ep V  \right ) \exp\left (\ep T-
{1\over {24}}\ep^3[T,[T,V]]\right ) \exp\left ({1\over 2} \ep V
\right ).
\label{newtb}
\ee
In this case  $\exp({1\over 2} \ep V)
\exp(\ep T) \exp({1\over 2} \ep V )$ has $e_{TTV}=1/12$ and
$e_{VTV}=1/24$ and propagator $\rho_{2B}$ corresponds to changing
$e_{TTV}$'s value to match that of $e_{VTV}$. The Takahashi-Imada
propagator (\ref{ti}) can also be written as
\be \rho_{TI}^\prime=
\exp\left ({1\over 2}\ep V + {1\over {48}}\ep^3[V,[T,V]] \right )
\exp\left (\ep T\right ) \exp\left ({1\over 2}\ep V + {1\over
{48}}\ep^3[V,[T,V]] \right ),
\label{tiprime}
\ee corresponding to
changing $e_{VTV}$'s value to match that of $e_{TTV}$. These are
the four fundamental correctable second order propagators with a
fourth order trace.

For the computation of the trace, it is unnecessary to know the corrector
explicitly. In other cases, such as symplectic corrector algorithms, one
may wish to apply the corrector occasionally to see the working of the
corrected fourth order propagator $\widetilde\rho$. We will give a detailed
derivation of correctors for propagators (\ref{ti})-(\ref{tiprime}),
cumulating in a set of four minimal correctors.
These minimal correctors with analytical coefficients have not
been previously described in
the literature\cite{wis96,mcl962,mar96,mar97,blan99,blan01}.

For the Takahashi-Imada propagator, we have $e_{TTV}=e_{VTV}=e_2$
with $e_2=-1/24$. From (\ref{gcor}), we see that a possible corrector is
$C=e_2\ep[T,V]$. This can be constructed in a straightforward manner
as suggested by Wisdom {\it et al.}\cite{wis96}. Since
\ba
B(v_1,t_1)&\equiv&
\exp(\ep v_1 V)\exp(\ep t_1 T)
\exp(-\ep v_1 V)\exp(-\ep t_1 T)\nonumber\\
&=&\exp\left(-v_1t_1\ep^2[T,V]-{1\over 2}t_1^2v_1\ep^3[T,[T,V]]
-{1\over 2}t_1v_1^2\ep^3[V,[T,V]]+O(\ep^4)\right), \label{ppmm}
\ea
by setting $v_1t_1=(1/48)$, the following product is a
workable corrector \be B(v_1,t_1)B(-v_1,-t_1) =\exp\left(-{1\over
24}\ep^2[T,V]+O(\ep^4)\right). \label{wiscor} \ee Note that it is
important to have the operator $V$ before $T$ to generate a
negative $e_2$ coefficient. However, without fully determining
both $v_1$ and $t_1$, this corrector clearly under-utilizes
$B(v_1,t_1)$. It requires eight operators, which is far from
optimal. We will show below that four is sufficient.

Let $H=T+V$ and $G=[T,V]$. Since
$H_A=H+\e_2\ep^2[H,G]$, we can see from (\ref{gcor}) that adding a
term $c_0H$ to $C$, will not affect the corrector term $\ep[C,T+V]$,
but such a term will generate unwanted third order terms
$c_0e_2\ep^3[H,[H,G]]$ from $\ep[C,H_A]$ and
${1\over 2}c_0e_2\ep^3[H,[G,H]]$ from ${1\over 2}\ep[C,[C,H_A]]$.
To cancel them, we must add another term $c_2\ep^2[H,G]$ to
the corrector such that $c_2={1\over 2}c_0e_2$. Thus the corrector
can have the more general form
\ba
\exp(\ep C)&=&\exp\left(c_0\ep H+\e_2\ep^2G
+{1\over 2}c_0e_2\ep^3[H,G]\right)+O(\ep^4),
\label{corcond}\\
&=&\exp(c_0\ep H)\exp(\e_2\ep^2G)+O(\ep^4),
\label{corsec}
\ea
where the second line follows from the fundamental Baker-Campbell-Hausdorff
formula, $\exp(A)\exp(B)=\exp(A+B+(1/2)[A,B]\cdots)$. To exploit
the use of the free parameter $c_0$, we can approximate
$\exp(c_0\ep H)$ by
\ba
&&\exp(\ep {c_0\over 2}V)\exp(\ep c_0T)\exp(\ep{c_0\over 2}V)\nonumber\\
&&=\exp\left(c_0\ep H+{1\over{12}}c_0^3\ep^3[T,[T,V]]
+{1\over{24}}c_0^3\ep^3[V,[T,V]]\right)+O(\ep^5),
\label{happ}
\ea
and the term $\exp(\e_2\ep^2G)$ by $B(v_1,t_1)$. We can now choose
$c_0,v_1,t_1$ such that $v_1t_1=1/24$ and the third order
terms in (\ref{happ}) exactly cancel the third order terms in (\ref{ppmm}):
${1\over 2}t_1^2v_1={1\over{12}}c_0^3$,
${1\over 2}t_1v_1^2={1\over{24}}c_0^3$. This gives
$c_0=1/(2\cdot 3^{1/6})$, $v_1=1/(4\sqrt{3})$ and $t_1=1/(2\sqrt{3})$. The
result is a corrector with six operators:
\be
S=\exp(\ep {c_0\over 2}V)\exp(\ep c_0T)
\exp(\ep ({c_0\over 2}+v_1)V)
\exp(\ep t_1 T)
\exp(-\ep v_1 V)\exp(-\ep t_1 T).
\label{sixop}
\ee
Since this corrector has made good use of all the parameters,
it is surprising that one can find a even shorter corrector.
Instead of $B(v_1,t_1)$,
consider just
\ba
&&\exp(\ep d_0 V)\exp(\ep d_0 T)\nonumber\\
&&=\exp\left(d_0\ep H-{1\over 2}d_0^2[T,V]
+{1\over{12}}d_0^3\ep^3[T,[T,V]]
-{1\over{12}}d_0^3\ep^3[V,[T,V]]\right)+O(\ep^4).
\ea
The corrector
\ba
S_{TI}
&=&\exp(\ep{c_0\over 2}V)\exp(\ep c_0T)
\exp(\ep ({c_0\over 2}+d_0)V)\exp(\ep d_0 T)\label{corti}\\
&=&\exp\biggl((c_0+d_0)\ep H+(-{1\over 2}d_0^2)\ep^2 G
+{1\over 2}(-{1\over 2}d_0^2)(c_0+d_0)[H,G]\nonumber\\
&&+{1\over{12}}(c_0^3+4d_0^3)\ep^3[T,[T,V]]
+{1\over{24}}(c_0^3+4d_0^3)\ep^3[V,[T,V]]\biggr) +O(\ep^4)
\nonumber
%\label{corfour}
\ea
will have the correct value for $e_2$ if we
take $d_0^2/2=1/24$, fixing $d_0=1/(2\sqrt{3})$. The corrector
will also be of the form (\ref{corcond}) after both commutators
have been eliminated by setting $c_0^3=-4 d_0^3$,
giving $c_0=-1/(2^{1/3}\sqrt{3})$.
This is the minimal corrector for the Takahashi-Imada propagator.

The corrector of the form (\ref{corcond}) is completely determined
by a single number $e_2$. Its sign dictates the order of the
$T$ and $V$ operators and its value fixes their coefficients.
For the alternative propagator $\rho_{2B}^{\,\prime}$ (\ref{newtbp})
with $e_2=-1/12$, its corrector is of the same form as (\ref{corti}),
but now with $d_0=1/\sqrt{6}$ and $c_0=-2^{1/6}/\sqrt{3}$.

For positive values of $e_2$, the corrector is of the form
\ba
S
&=&\exp(\ep{c_0\over 2}T)\exp(\ep c_0V)
\exp(\ep ({c_0\over 2}+d_0)T)\exp(\ep d_0 V)\label{cornew}\\
&=&\exp\biggl((c_0+d_0)\ep H+({1\over 2}d_0^2)\ep^2 G
+{1\over 2}({1\over 2}d_0^2)(c_0+d_0)[H,G]\nonumber\\
&&-{1\over{24}}(c_0^3+4d_0^3)\ep^3[T,[T,V]]
-{1\over{12}}(c_0^3+4d_0^3)\ep^3[V,[T,V]]\biggr) +O(\ep^4).
\nonumber
%\label{corfive}
\ea
Propagator $\rho_{2B}$ is dual to the $TI$ propagator with $e_2=1/24$.
Its corrector is of the form (\ref{cornew}) but with same coefficients
$d_0=1/(2\sqrt{3})$ and $c_0=-1/(2^{1/3}\sqrt{3})$.
The $\rho_{TI}^{\,\prime}$ propagator (\ref{tiprime}) with $e_2=1/12$
is dual to $\rho_{2B}^{\,\prime}$. Its corrector is of the form
(\ref{cornew}) with $d_0=1/\sqrt{6}$ and $c_0=-2^{1/6}/\sqrt{3}$.
These compact correctors are fitting companions to their equally
compact propagators.

\section{Conclusions}

     In this work, we proved a fundamental result
on the correctability of forward time step propagators. We show
that if $\rho=\e^{\ep(T+V)}$ were to be approximated by the product
form (\ref{arho}), then no product form with positive coefficients
$\{t_i\}$ is correctable beyond second order. Whereas a conventional
higher order propagator requires its error terms to vanish,
a correctable propagator only require its error terms to satisfy
the correctability condition. The latter requirement seemed far
less stringent. A surprising element of this work is that, this is
not the case. For symmetric decomposition with positive $\{t_i\}$, the two
second order error coefficients cannot both vanish because, they can
never be equal! The correctability requirement itself is stringent enough.
This proof of non-correctability generalizes the previous work of
Sheng\cite{sheng} and Suzuki\cite{suzukinogo}.

    From knowing correctability requirement, we derived
systematically the four forward correctable second order
propagators and their minimal correctors. These minimal correctors
follow from a more general form (\ref{corsec}) of the corrector
with free parameters. Much of the existing literature on
symplectic corrector is rather opaque, concerned only with how to
satisfy ``order conditions" numerically\cite{blan99,blan01}. This
work suggests that a more analytical approach is possible.

The Takahashi-Imada type of propagators considered here are unique
in that they are the only known second-order, forward-time-step
propagators with a fourth order trace. If one is willing to
evaluate the potential at least twice, then with the inclusion of
$[V,[T,V]]$, one can make both error coefficients $e_{TTV}$ and
$e_{VTV}$ vanish\cite{suzfour,chin}. The result is a whole family
of positive time step fourth order
propagators\cite{chinchen02,ome02,ome03,chinchen03} with a fourth
order trace. While this class of forward decomposition algorithms
is indispensable for solving time-irreversible
problems\cite{fchinl,fchinm,auer,ochin,jang}, they are less
interesting from the point of view of calculating the trace. For
correctable propagators, their key attraction is that one can
obtain a higher order trace without using a higher order
propagator. Methods and results of this work can be used to study
ways of correcting these fourth order propagators to higher
orders.

\begin{acknowledgments}

I wish to thank J. Boronat and J. Casulleras, for their invitation to
lecture in Barcelona during the summer of 2003 which initiated
this work, E. Krotscheck, for his interest and hospitality at Linz,
H. Forbert, for discussing the correctability requirement, and G. Chen,
on the use of constrained minimization. This work was supported,
in part, by a National Science Foundation grant, No. DMS-0310580.

\end{acknowledgments}
%%%%%%%%%%%%%%%%%%%%%%%%%%%%%%%%%%%%%%%%%%%%%%%%%%%%%%%%%%%%%%%%%%%%%%%%%%%%%%%%%
%\newpage
\bigskip
\bigskip
\centerline{REFERENCES}
%\vspace{.3 truein}

%\bibliography {dmts}

\begin{thebibliography}{10}

\bibitem{sheng}Q. Sheng, IMA J. Num. Anaysis, {\bf 9},
              199 (1989).
\bibitem{suzukinogo}M. Suzuki, J. Math. Phys. {\bf 32}, 400 (1991).
\bibitem{goldman}D. Goldman and T. J. Kaper, SIAM J. Numer. Anal.,{ \bf 33},
                   349 (1996).
\bibitem{mcl91}R. I. McLachlan and P. Atela,
         Nonlinearity, {\bf 5}, 542 (1991).
\bibitem{mcl95}R. I. McLachlan, SIAM J. Sci. Comput. {\bf 16}, 151 (1995).
\bibitem{kos96}P. V. Koseleff, in
          {\it Integration algorithms and classical mechanics}, Fields Inst. Commun.,
          10, Amer. Math. Soc., Providence, RI, P.103, (1996).
\bibitem{mcl02} R. I. McLachlan and G. R. W. Quispel, Acta Numerica,
          {\bf 11}, 241 (2002).
\bibitem{suzfour}M. Suzuki, {\it Computer Simulation Studies in
            Condensed Matter Physics VIII},
           eds, D. Landau, K. Mon and H. Shuttler (Springler, Berlin, 1996).
\bibitem{chin} S.A. Chin, Physics Letters {\bf A} 226, 344 (1997).
\bibitem{fchinl}H. A. Forbert and S. A. Chin,
                Phys. Rev. {\bf E 63}, 016703 (2001).
\bibitem{fchinm}H. A. Forbert and S. A. Chin,
                Phys. Rev. {\bf B 63}, 144518 (2001).
\bibitem{auer}J. Auer, E. Krotscheck, and S. A. Chin,
                J. Chem. Phys. {\bf 115}, 6841 (2001).
\bibitem{ochin}O. Ciftja and S. A. Chin, Phys. Rev.
               {\bf B 68}, 134510 (2003).
\bibitem{jang} S. Jang, S.Jang and G. A. Voth,
              J. Chem. Phys. {\bf 115} 7832, (2001).
\bibitem{kalos} M. Kalos, D. Levesque, and L. Verlet,
                Phys. Rev. {\bf A 9}, 2178 (1974).
\bibitem{chin90}S. A. Chin, Phys. Rev. {\bf A 42}, 6991 (1990).
\bibitem{klein} H. Kleinert, {\it Path Integrals in Quantum Mechanics,
                 Statistics and Polymer Physics}, World Scientific,
                 Singapore, 1990.
\bibitem{wis96}J. Wisdom, M. Holman, AND J. Touma, ``Symplectic correctors",
          in {\it Integration Algorithms and Classical Mechanics},
          J. E. Marsden, G. W. Patrick, and W. F. Shadwick, eds., American Mathematical
          Society, Providence, RI, 1996.
\bibitem{mcl962}R. I. McLachan, ``More on symplectic correctors", in {\it Integration Algorithms
         and Classical Mechanics}, J. E. Marsden, G. W. Patrick, and W. F. Shadwick, eds.,
         American Mathematical Society, Providence, RI, 1996.
\bibitem{mar96}M. A. Lopez-Marcos, J. M. Sanz-Serna, and R. D. Skeel,
            in Numerical Analysis 1995,
            D. F. Griffiths and G. A. Watson,
            eds., Longman, Harlow, UK, 1996, pp. 107-122.
\bibitem{mar97}M. A. Lopez-Marcos, J. M. Sanz-Serna, and R. D. Skeel,
           SIAM J. Sci. Comput., {\bf 18} 223, (1997).
\bibitem{blan99}S. Blanes, F. Casas, and J. Ros,
        Siam J. Sci. Comput., {\bf 21}, 711 (1999).
\bibitem{blan01} S. Blanes, App. Numer. Math {\bf 37}, 289 (2001).
\bibitem{suz95}M. Suzuki, Phys. Lett. {\bf A 201}, 425 (1995).
\bibitem{ti}M. Takahashi and M. Imada,
          J. Phys. Soc. Jpn {\bf 53}, 3765 (1984).
\bibitem{li} X.-P. Li, J. Q. Broughton,
           J. Chem. Phys., {\bf 86}, 5094 (1987)
\bibitem{chinchen02}S. A. Chin and C. R. Chen,
                J. Chem. Phys. {\bf 117}, 1409 (2002).
\bibitem{ome02}I. P. Omelyan, I. M. Mryglod and R. Folk,
               Phys. Rev. {\bf E66}, 026701 (2002).
\bibitem{ome03}I. P. Omelyan, I. M. Mryglod and R. Folk,
               Comput. Phys. Commun. {\bf 151} 272 (2003)
\bibitem{chinchen03}S. A. Chin, and C. R. Chen,
``Forward Symplectic Integrators for Solving Gravitational
   Few-Body Problems", arXiv, astro-ph/0304223.

\end{thebibliography}
%\bibliographystyle{revtex4}

%%%%%%%%%%%%%%%%%%%%%%%%%%%%%%%%%%%%%%%%%%%%%%%%%%%%%%%
\end{document}